\documentclass[10pt]{article}
\usepackage{times}
\usepackage{multicol,setspace}
\usepackage{hyperref}
\usepackage{fontawesome}
\usepackage[fleqn]{amsmath}
\usepackage{amssymb}
\usepackage{authblk}
\usepackage{fancyhdr}
\usepackage{picins}
\usepackage{lastpage}
\usepackage{geometry}
\usepackage{orcidlink}
\usepackage{graphicx,caption}
\usepackage{nomencl}
\usepackage[numbers,longnamesfirst]{natbib}
\usepackage[X2,OT1]{fontenc}
\usepackage[document]{ragged2e}
\usepackage{setspace}
\usepackage[main=english,russian]{babel}

\hypersetup{
  colorlinks = true,
  linkcolor=red,
  linkbordercolor= {0 0 1}}
  
\providecommand{\keywords}[1]
{
  \textbf{Keywords:} #1
}

\makeatletter
\newenvironment{tablehere}
  {\def\@captype{table}}
  {}

\newenvironment{figurehere}
  {\def\@captype{figure}}
  {}
\makeatother

\begin{document}
\justifying
\makenomenclature

\title{\vspace{-2.5em} \normalsize IAC-22-E3-1 \vspace{1em} \\ \normalsize \textbf{Vela: A Data-Driven Proposal for Joint Collaboration in Space Exploration}}

\author[1\textsection]{\textbf{\normalsize{Holly M. Dinkel}}\orcidlink{0000-0002-7510-2066}}
\author[2\textsection]{\textbf{\normalsize{Jason K. Cornelius}}}
\affil[1]{\normalsize{\textit{Department of Aerospace Engineering at the University of Illinois at Urbana-Champaign, Urbana, IL, USA}}}
\affil[2]{\normalsize{\textit{Department of Aerospace Engineering at the Pennsylvania State University, State College, PA, USA}
$^1$\texttt{hdinkel2@illinois.edu}, $^2$\texttt{joc5693@psu.edu}}}
\date{\vspace{-5ex}}
\maketitle

\begingroup\renewcommand\thefootnote{\textsection}
\footnotetext{Equal contribution}
\endgroup

\vspace{-2em}

\normalsize \begin{abstract}

\normalsize
The UN Office of Outer Space Affairs identifies synergy of space development activities and international cooperation through data and infrastructure sharing in their Sustainable Development Goal 17 (SDG17). Current multilateral space exploration paradigms, however, are divided between the Artemis and the Roscosmos-CNSA programs to return to the moon and establish permanent human settlements. As space agencies work to expand human presence in space, economic resource consolidation in pursuit of technologically ambitious space expeditions is the most sensible path to accomplish SDG17. This paper compiles a budget dataset for the top five federally-funded space agencies: CNSA, ESA, JAXA, NASA, and Roscosmos. Using time-series econometric anslysis methods in STATA, this work analyzes each agency's economic contributions toward space exploration. The dataset results are used to propose a multinational space mission, Vela, for the development of an orbiting space station around Mars in the late 2030s. Distribution of economic resources and technological capabilities by the respective space programs are proposed to ensure programmatic redundancy and increase the odds of success on the given timeline. 

\end{abstract}

\begin{multicols}{2}
\normalsize
\renewcommand{\nomname}{\large{Acronyms and Abbreviations}}
\small
\nomenclature{\textbf{CNSA}}{China National Space Administration}
\nomenclature{\textbf{ESA}}{European Space Agency}
\nomenclature{\textbf{GDP}}{Gross Domestic Product}
\nomenclature{\textbf{ISS}}{International Space Station}
\nomenclature{\textbf{JAXA}}{Japanese Aerospace Exploration Agency}
\nomenclature{\textbf{LEO}}{Low Earth Orbit}
\nomenclature{\textbf{NASA}}{National Aeronautics and Space Administration}
\nomenclature{\textbf{P/L}}{Payload}
\nomenclature{\textbf{R\&D}}{Research and Development}
\nomenclature{\textbf{SDG}}{Sustainable Development Goal}
\nomenclature{\textbf{SLS}}{Space Launch System}
\nomenclature{\textbf{USD}}{United States Dollars}
\nomenclature{\textbf{VECM}}{Vector Error Correction Mechanism}
\nomenclature{\textbf{\$B}}{Billions of United States Dollars}
\printnomenclature
\normalsize
\subsection{Introduction}

The question of whether humanity can become an interplanetary species is fundamental to human curiosity and its answer is important for basic human survival. Establishing a human presence at Mars is essential to answering this question, but it competes for resources with objectives aimed at directly improving life on earth. Efforts to reduce direct costs associated with human space exploration and to share costs among international partners will be key to establishing human space exploration as financially and technologically viable.

Earth observation and satellite communications are the technologies with the most alignment to the Sustainable Development Goals (SDGs) of the UN \cite{Baumgart_2021}. The commercialization of Low Earth Orbit (LEO) lead to a factor of 20 reduction in launch costs between the 1990s and 2010 when the SpaceX Falcon 9 rocket first launched, increasing access to space through reduced launch costs and greater availability of launch vehicles \cite{Jones_2018}. This alignment is due in part to the high technology readiness level of the sensors and communication relays in these satellites as well as from routine launching to Earth orbits during the 2000s. National space program objectives are pivoting from focus on access to \textit{space} to access to \textit{deep space} (the Moon and Mars), with international efforts to establish human-inhabited deep space outposts underway \cite{Artemis_2020,MoonBase}.

\subsubsection{Space Capabilities Around the World}

This article investigates the technological and financial capability of governmental space programs to cooperate on sending humans to Mars. China, the EU, Japan, Russia, and the USA were selected for this study based on three criteria:

\begin{enumerate}
    \item All five space agencies have experience manufacturing and launching heavy-lift launch vehicles capable of reliably performing deep space missions.
    \item All five space agencies successfully developed and launched orbital space station habitats (either modules or full space stations) capable of supporting human life in space.
    \item All five space agencies have experience with launching robotic missions to Mars.
\end{enumerate}

A super heavy-lift launch vehicle is capable of lifting 50,000 kg to LEO \cite{NASA_heavylift}. Table \ref{tab:launchcapability} summarizes the current and planned super heavy-lift launch vehicles which are shown in Figure \ref{fig:heavy_lift_lvs}. Notably, China, Russia, and USA are the only nations developing this capability. Table \ref{tab:habitatcapability} summarizes the space station habitats produced by each government and their mass at launch. Importantly, these data show that a space station module comparable in mass to extant modules on the ISS launched to deep space will require a launch on a super heavy-lift launch vehicle. The cost to produce an expandable module for future space stations is expected to be \$300M, while the cost to produce an ISS-heritage module is estimated to be \$250M \cite{Crane_2017}. Table \ref{tab:marscapability} summarizes launches each agency has sent to Mars, where launch windows occur every $\sim$1.6 years during the closest orbital approach between Earth and Mars \cite{PlanetarySociety}. In Table \ref{tab:launchcapability} and Table \ref{tab:marscapability}, costs are reported as the cost-per-launch for each launch vehicle. These costs excludes mission-related costs such as the cost of spacecraft manufacture.

\begin{center}
\begin{tablehere}
\centering
\captionsetup{type=table}
\caption{Current and Planned Heavy-Lift Launch Vehicles}
\footnotesize
\begin{tabular}{ccccc}
\hline \hline
Status$^*$ & Vehicle & P/L to LEO (kg) & Cost (\$M)$^\dagger$ \\
\hline
A & Falcon Heavy & 63,800 & \$97.00 \\
A & SLS Block 1 & 95,000 & \$2,800.00 \\
P & Starship & 100,000 & N/A \\
P & Yenisei & 103,000 & N/A \\
P & SLS Block 1B & 105,000 & N/A \\
P & SLS Block 2 & 130,000 & N/A \\
P & Long March 9 & 150,000 & N/A  \\
\hline \hline
\end{tabular}
\newline
\footnotesize{\begin{tabular}{cc}$^*$A: Available, P: Planned & $^\dagger$N/A: Not Available \end{tabular}}
\label{tab:launchcapability}
\end{tablehere}
\end{center}

\begin{figurehere}
\centering
\captionsetup{type=figure}
\includegraphics[width=0.48\textwidth]{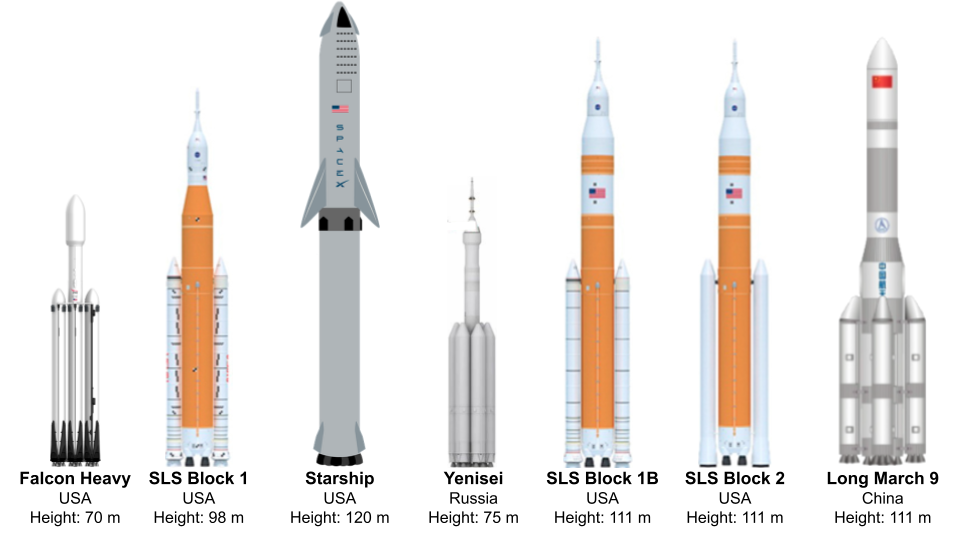}
\caption{Super heavy-lift launch vehicles capable of lifting at least 50,000 kg P/L to LEO (current and planned).}
\label{fig:heavy_lift_lvs}
\end{figurehere}

\begin{center}
\begin{tablehere}
\centering
\captionsetup{type=table}
\caption{Current Capability in Human Habitat Development for LEO}
\footnotesize
\begin{tabular}{cccccc}
\hline \hline
Year & Government & Station & Habitat Module & Mass (kg) \\
\hline
2022 & China & Tiangong & Wentian & 23,200 \\
2021 & China & Tiangong & Tianhe & 22,600 \\
2021 & Russia & ISS & Nauka & 20,357 & \\
2009 & Japan & ISS & Kibo & 15,900 \\
2008 & EU & ISS & Columbus & 10,300 \\
2007 & USA & ISS & Harmony & 14,300 \\
2001 & USA & ISS & Destiny & 14,515 \\
2000 & Russia & ISS & Zvezda & 20,320 \\
1998 & USA & ISS & Unity & 11,612 \\
1998 & Russia & ISS & Zarya & 19,323 \\
\hline \hline
\end{tabular}
\label{tab:habitatcapability}
\end{tablehere}
\end{center}

\begin{center}
\begin{table*}
\centering
\caption{Current Launch Capability to Mars}
\footnotesize
\begin{tabular}{cccccc}
\hline \hline
Year & Vehicle & Government & Mission (P/L Type$^\dagger$) & P/L Mass (kg) & Cost (\$M)$^\mathsection$ \\
\hline
2020 & H-IIA & Japan & Emirates Mars Mission (O) & 1,350 & \$90.00 \\
2020 & Long March 5 & China & Tianwen-1 (L + O) & 5,000 & \$150.00$^\ddagger$\\
2020 & Atlas V & USA & Mars 2020 (L) & 3,649 & \$243.00\\
2020 & Atlas V & USA & InSight (L) & 721 & \$163.40 \\
2016 & Proton-M / Briz-M & Russia$^*$ & ExoMars (L + O) & 4,332 & N/A \\
2011 & Atlas V 541 & USA & Mars Science Laboratory (L) & 3,839 & N/A \\
2007 & Delta II 7925 & USA & Phoenix (L) & 670 & \$86.20\\
2005 & Atlas V 401 & USA & Mars Reconnaissance Orbiter (O) & 2,180 & \$90.00 \\
2003 & Soyuz-FG / Fregat & Russia$^*$ & Mars Express (L + O) & 1,189 & N/A \\
2003 & Delta II 7925 & USA & Mars Exploration Rover Mission - Spirit (L + O) & 1,063 & N/A \\
2003 & Delta II 7925H & USA & Mars Exploration Rover Mission - Opportunity (L + O) & 1,063 & N/A \\
2001 & Delta II 7925-9.5 & USA & 2001 Mars Odyssey (O) & 758 & \$53.90\\
1999 & Delta II 7425-9.5 & USA & Mars Polar Lander (L) & 290 & \$90.70\\
1998 & M-V & Japan & Nozomi (O) & 258 & \$70.00\\
1998 & Delta II 7425-9.5 & USA & Mars Climate Orbiter (O) & 638 & \$90.70\\
\hline \hline
\end{tabular}
\footnotesize{\begin{tabular}{ccccc} Note: Country, Payload Mass, and Cost are provided for launch. & $^*$Payload: EU & $^\dagger$L: Lander, O: Orbiter & $^\ddagger$Estimated & $^\mathsection$N/A: Not Available \end{tabular}}
\label{tab:marscapability}
\end{table*}
\end{center}

\subsubsection{Why Cooperate in Space Exploration?}

Historically, the financial risks associated with space development motivated cooperation in space. Spreading costs among partners decreases the total percentage of program costs shouldered by any individual nation; by increasing access to the program, the total utility of the endeavor increases. International cooperation also increases political stability. International partnership guards programs from cancellation by national administrations unwilling to renege on agreements to protect the national diplomatic reputation. Successful international partnerships improve a nation's reliability, and unilateral departure from an ongoing partnership may signal unpredictability to future candidate partners \cite{Broniatowski_2006}.

Perhaps the most prominent example of successful international collaboration in pursuit of shared objectives in space is the construction of the International Space Station (ISS). The ISS is technologically complex, with a total cost of construction and operation estimated to be \$150B. With costs and resources shared among partner space agencies, it has enabled nearly 3,000 scientific experiments and the development of new science and technology on earth \cite{NASA_2022}. The challenges associated with the construction of the ISS caution any actor pursuing future space habitat construction, as "final configuration of the ISS cost more, took longer to complete, and is less capable than planners envisioned" \cite{NASA_2013}. The financial and technological redundancy offered through international partnership are arguably the cornerstones of the viability of the program. 

There are two government-led programs in development for reactivating human deeps space exploration. NASA is establishing the Artemis program to develop an orbiting deep space gateway around the moon with ESA, the Canadian Space Agency, and JAXA. CNSA and Roscosmos have jointly announced a partnered program to establish a human settlement on the surface of the moon. As these programs evolve, they will provide greater insight into efficient international resource allocation for human exploration of deep space.
\subsection{Objectives and Methodology}
\label{sec:methods}
\subsubsection{Objectives}
\label{sec:objectives}
This work presents a quantitative econometric analysis of macroeconomic data for the national space programs of China, the EU, Japan, Russia, and the USA. The data and econometric models highlight the current funding status of each agency and the trends that will impact future funding. The most recent space program budget data are used to propose a financial cooperation scenario for Vela: a human missions to Mars. The objective of this work is to design a multinational Mars mission with contributions from each space agency based on current or planned technological and financial infrastructure. The primary research question addressed in this work is: \textbf{how might space agencies contribute technological and financial resources toward a shared objective of establishing a human presence at Mars?} To explore this question, this work sets three objectives:

\begin{enumerate}
    \item Investigate human spaceflight, Mars exploration, and launch capabilities for each government. These capabilities may indicate the readiness for a government to contribute a specific technology to a human Mars mission.
    \item Investigate the time series budget data for each government space agency and determine factors with which the data are correlated. Identifying correlation relationships between a national space budget and other national economic circumstances provides insight to the sensitivity of space funding levels to macroeconomic conditions.
    \item Plan a human mission to Mars based on available spaceflight capabilities, estimated direct costs associated with spaceflight capabilities, and the current budget data for each government space agency. The overall utility and sustainability of space exploration programs increases when costs are shared among partnering organizations, and multi-national risk-sharing for a human mission to Mars is necessary for its achievement.
\end{enumerate}

The authors hypothesize China, Russia, and the USA will be the primary providers of human-rated heavy-lift launch vehicles capable of transporting crew to Mars. This is based on both current and planned launch capabilities stated by the respective nations. Based on the current funding levels and allocated annual budgets, the authors hypothesize China and the USA will be the primary providers of heavy-lift launch vehicles for transporting cargo to Mars. Russia and the USA both have over fifty years of experience with launching people into space and are good candidates for leading crew training and human launch efforts \cite{GLEX_2021}. JAXA and ESA each manufactured science laboratory modules for the ISS (Kibo and Columbus, respectively) and are experienced with designing functional human space habitats. Incorporating this information about flight heritage and demonstrated strengths during during planning for a cooperative human space mission will likely increase the probability of a successful mission.

\subsubsection{Methodology}
\label{sec:methodology}

\begin{figure*}[t]
\centering
\includegraphics[width=0.9\textwidth]{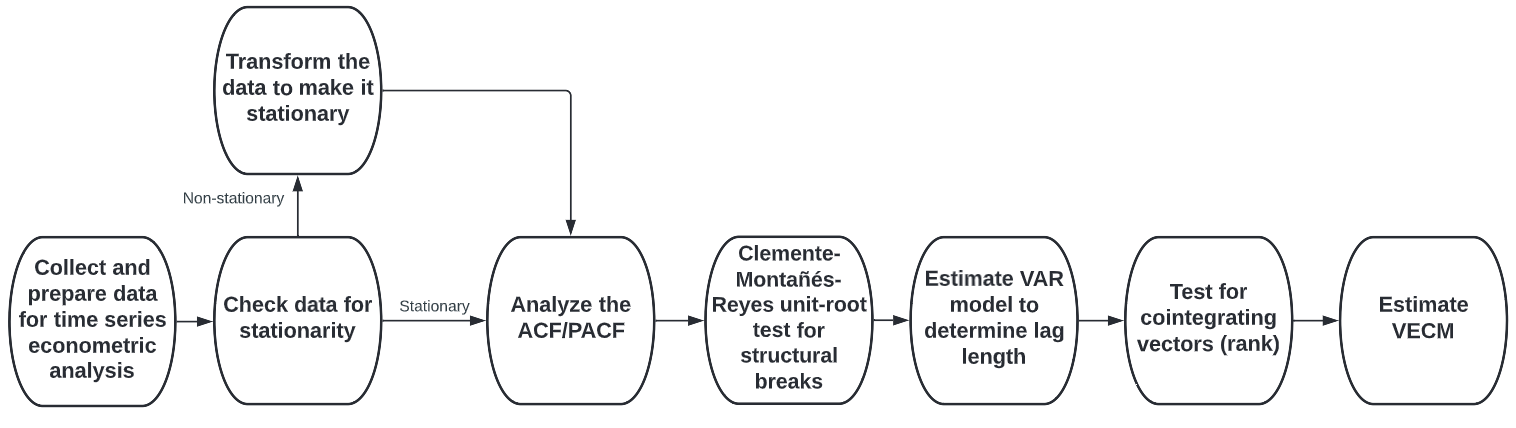}
\caption{The procedure for performing econometric analysis of macroeconomic data includes preparing the data for analysis, analyzing the data for autocorrelation and structural breaks, determining the lag lengths, testing for cointegrating vectors, and estimating the VECM model.}
\label{fig:econometrics_procedure}
\end{figure*}

The authors compiled a dataset comprising annual macroeconomic data for China, the EU, Japan, Russia, and the USA from 1998--2020. This dataset was used to model funding levels for each of the national space agencies: CNSA (China), ESA (EU), JAXA (Japan), Roscosmos (Russia), and NASA (USA). The data categories include national space agency budget, Gross Domestic Product in 2020 USD, country population, researchers per million people, military spending as a percentage of GDP, education spending as a percentage of GDP, and science research and development (R\&D) spending as a percentage of GDP \cite{NASABUDGET,WorldBank,Statista,OECD_2007,GlobalSecurity,ESA_2002,ESA_2006,ESA_2008,ESA_2020,Roscosmos_1998,Roscosmos_2011,JAXA}. For any missing data entries, the missing value was linearly interpolated between neighboring (in time) data points to complete the dataset. 

Once the data set was complete, econometric analysis was used to identify causal relationships between the dependent variable (i.e., space budget) and the remaining independent variables. Econometric analysis is a statistical approach used to quantify between-variables effects in multivariate systems over a period of time. The Johansen Vector Error-Correction Mechanism (VECM) approach was used to identify correlation between the variables \cite{Fedderke_1999,Fedderke_2005,Fedderke_2008,GLEX_2021}. The Johansen VECM is a robust time-series analysis technique used for analysis of non-stationary and potentially co-integrating data. The Johansen VECM identifies co-integrating vectors that predict dependent variables as functions of multiple independent variables changing over time; the Johansen VECM estimation procedure is standard practice in econometric analysis so its discussion here will be brief \cite{Johansen_1991}. To estimate the value of a dependent variable, $z$, a set of $k$ independent variables are required within which there may be $r$ co-integrating relationships such that $0 \leq r \leq k-1$. 
See Equation \ref{eq: vecm}.

\begin{equation}
\label{eq: vecm}
    \mathrm{\Delta} z_t = \sum_{i=1}^{k-1} \mathrm{\Gamma}_i \mathrm{\Delta} z_{t-i} + \mathrm{\Pi} z_{t-k+1} + \mu + \delta_t
\end{equation}

\noindent In this formulation, $\mu$ is a set of deterministic components and $\delta$ is a Gaussian error term. The existence of $r$ cointegrating relationships results in the hypothesis
\begin{equation}
\label{eq: hypothesis}
\begin{matrix}
H_1(r): & \mathrm{\Pi} = \alpha \beta',
\end{matrix}
\end{equation}

\noindent where $\mathrm{\Pi} \in \mathbb{R}^{p\times p}$ is the long-run impact matrix and $\alpha, \beta \in \mathbb{R}^{p \times r}$ are full-rank weights and co-integration vectors, respectively. 

To state the result more succinctly, consider a model of the form 
\begin{equation}
\label{eq:model}
\ln{SB} = \begin{bmatrix} \beta_1 \\ \beta_2 \\ \beta_3 \\ \beta_4 \\ \beta_5 \end{bmatrix}^\intercal \begin{bmatrix} \ln{GPC} \\ \ln{RD} \\ \ln{MD} \\ \ln{ED} \\ \ln{SD} \end{bmatrix} + C,
\end{equation}

\noindent where
\begin{gather*}
SB = \text{National Space Agency Budget (\$B);} \\
GPC = \text{Gross Domestic Product per Capita (2020 USD);} \\
RD = \text{Researchers per Million People;} \\
MD = \text{Military Spending (\% GDP);} \\
ED = \text{Education Spending (\% GDP);} \\
SD = \text{Science R\&D Spending (\% GDP);} \\
\end{gather*} 

\noindent and the values of ${\beta}$ represent the coefficients in front of each independent variable and $C$ is the y-intercept value for the equation.

The process used in this study to carry out the econometric analysis from data collection through estimation of the VECM is summarized in Figure \ref{fig:econometrics_procedure}. First, data for the dependent variable ($SB$) and independent variables ($GPC$, $RD$, $MD$, $ED$, $SD$) were recorded annually. The aforementioned VECM estimation process was completed using the Stata/BE 17.0 data analysis software \cite{stata}. Tests for stationarity were conducted and the variables were then pre-processed into levels through a natural logarithm transform. This was required to make the data stationary, which is necessary for the VECM analysis process. 

The emphasis while developing the VECM models was to obtain a single co-integrating equation. As such, some model specifications were required to omit one or more of the independent variables due to high levels of auto-correlation in some of the data sets. This basically means that two independent variables can potentially be controlling for the same input into the space budget estimation, at which point more than one co-integrating equation would be required. For simplicity, the authors decided to omit one of the variables in this scenario and arrive at a model with a single co-integrating equation. Various specifications with different combinations of the independent variables and the model lag length were created. 

The accuracy of all resulting model specifications were then considered by comparing the ${\chi}^2$, Akaike Information Criterion (AIC), Bayesian Information Criterion (BIC), and Log Likelihood values for each model. The statistical significance of the resulting coefficients in each model were also considered. Tests for stability were also conducted to ensure all eigenvalue roots fall within the unit circle. As previously mentioned, the model specifications were targeting a single co-integrating vector and a lag length of one or two was used based on the results of the appropriate testing. This means that the variable values of the previous one to two time steps are used along with those of the current time step to estimate the dependent variable's current (or future) value. This approach of forcing a single co-integrating vector lead to multiple specifications for some of the datasets, which are all used to provide insight on the data in the Results section.
\subsection{Results and Limitations}
\label{sec:results}
\subsubsection{The Model}

\begin{table*}
\centering
\caption{Variable Correlation with National Space Budgets }
\includegraphics[width=0.75\textwidth]{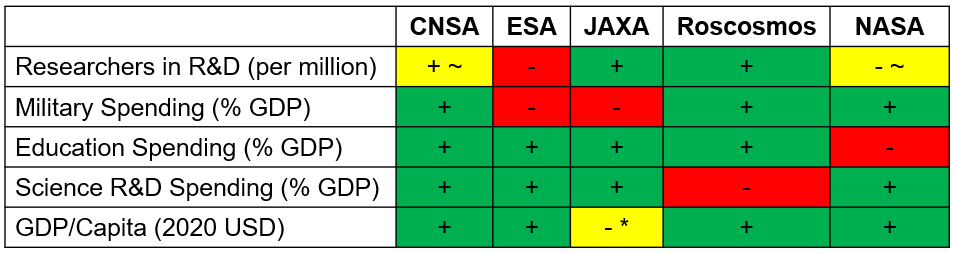}
\footnotesize{\begin{tabular}{c}$^*$Statistically insignificant correlation at 5\% Confidence Interval \\ $^\sim$Conflicting results from two different model specifications, value with larger Z-score used \end{tabular}}
\label{tab:trends}
\end{table*}

The econometric analysis applied the Johansen VECM to the time-series macroeconomic data of each government. The VECM analysis was used to predict the space budget of each country as a function of national GDP per capita, number of researchers per million people, military spending, education spending, and science R\&D spending. The data used to model the budgets of CNSA, ESA, JAXA, Roscosmos, and NASA exhibited multiple co-integrating relationships as determined by the VECRANK predictor in Stata. This necessitated multiple model specifications to fully capture the trends between all independent variables and the resulting space budgets for these agencies. The most meaningful models from the VECM analysis for each space agency, selected subjectively by the authors, are provided in Equations \ref{eq:china}--\ref{eq:usa}, and the complete set of models is provided in Appendix B. These models can be used to estimate the budget of each space agency as a function of the independent variables. The models capture the sensitivity of each space agency's budget to the independent variables. Correlating relationships between variables is valuable for understanding the implications of current events and policies on the dependent variable. The correlation between each independent variable and the dependent variable can be either positive, negative, or ambiguous, and they are summarized in Table \ref{tab:trends} for the models generated in this work. 

\begin{flalign}
\label{eq:china}
&\textbf{CNSA: } \ln{SB} = \begin{bmatrix} 4.9 \\ -4.6 \\ 33.4 \\ 0 \\ 0 \end{bmatrix}^\intercal \begin{bmatrix} \ln{GPC} \\ \ln{RD} \\ \ln{MD} \\ \ln{ED} \\ \ln{SD} \end{bmatrix} - 29.3&
\end{flalign}

\begin{flalign}
\label{eq:eu}
&\textbf{ESA: } \ln{SB} = \begin{bmatrix} 0.1 \\ -0.1 \\ -0.2 \\ 0.3 \\ 2.8 \end{bmatrix}^\intercal \begin{bmatrix} \ln{GPC} \\ \ln{RD} \\ \ln{MD} \\ \ln{ED} \\ \ln{SD}\end{bmatrix} - 1.3&
\end{flalign}

\begin{flalign}
\label{eq:japan}
&\textbf{JAXA: } \ln{SB} = \begin{bmatrix} -0.8 \\ 24.9 \\ -19.3 \\ 13.6 \\ 4.0 \end{bmatrix}^\intercal \begin{bmatrix} \ln{GPC} \\ \ln{RD} \\ \ln{MD} \\ \ln{ED} \\ \ln{SD} \end{bmatrix} - 225.7& \end{flalign}

\begin{flalign}
\label{eq:russia}
&\textbf{Roscosmos: } \ln{SB} = \begin{bmatrix} 0 \\ 0 \\ 6.4 \\ 5.1 \\ -5.5 \end{bmatrix}^\intercal \begin{bmatrix} \ln{GPC} \\ \ln{RD} \\ \ln{MD} \\ \ln{ED} \\ \ln{SD} \end{bmatrix} - 14.3&
\end{flalign}

\begin{flalign}
\label{eq:usa}
&\textbf{NASA: } \ln{SB} = \begin{bmatrix} 0.7 \\ -0.8 \\ 0.2 \\ 0 \\ 1.7 \end{bmatrix}^\intercal \begin{bmatrix} \ln{GPC} \\ \ln{RD} \\ \ln{MD} \\ \ln{ED} \\ \ln{SD} \end{bmatrix} + 0.3&
\end{flalign}

\subsubsection{International Cooperation for Vela}

\begin{figure*}
\centering
\captionsetup{type=figure}
\includegraphics[width=0.75\textwidth]{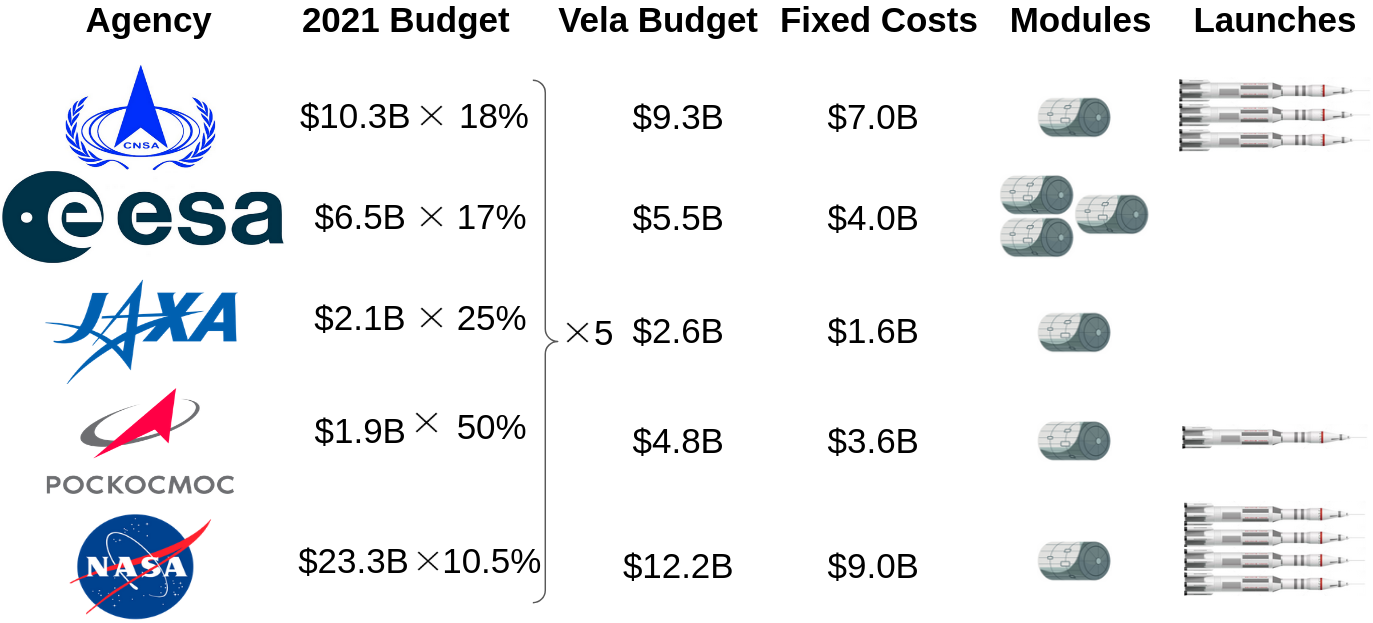}
\caption{Cost sharing among space agencies facilitates ambitious Mars missions.}
\label{fig:optimization}
\end{figure*}

The budget of a space agency is the keystone enabling its programs and projects. When agencies cooperate and contribute resources and capabilities toward a larger mission, they are able to pursue ambitious projects which would be challenging if not impossible for any individual agency to accomplish alone. This section uses the current budget for each space agency to propose a collaborative human exploration mission to Mars, named Vela. This scenario envisions development of a Mars-orbiting space station comprising several pressurized modules, a crew, and the launch infrastructure enabling its construction. 

Development costs of space station modules were estimated at \$300M, consistent with development costs of modules built for the ISS. It assumes the mass of each module is 15,000-20,000 kg, which is in line with the mass of modules for existing space habitats as reported in Table \ref{tab:habitatcapability}. This meant the modules must be carried on a super heavy-lift launch vehicle. The mission analysis assumes the available super heavy-lift launch vehicles will be manufactured in and launched from China, Russia, and the USA based on the current and planned launch vehicles reported in Table \ref{tab:launchcapability}. Without real cost data, the mission plan assumes the cost-per-launch for Chinese and Russian super heavy-lift launch vehicles is the same as the cost of launch for one SLS rocket based on comparable total program development costs across vehicles\footnote{Because launch cost estimates for the Long March 9 are unavailable, its cost-per-launch is not discounted due to its reusability. Future mission analysis scenarios should account for the financial sustainability of reusable (as opposed to expendable) launch vehicles as their cost estimates become available.}. This could be a conservative estimate due to the NASA programmatic cost overruns of the SLS program \cite{Ansar_2022}.

The proposed Vela mission analysis is presented in Figure \ref{fig:optimization}. Each agency's contributions to the Vela mission are summarized in its respective row. The 2021 space budget for each agency is reported in USD. Since NASA, CNSA, and Roscosmos are currently developing super heavy-lift launch capability, the number of launches from each agency was divided to reflect their differences in budget allocations. Furthermore, the number of modules contributed from each agency was distributed across all agencies while allotting a higher responsibility to the ESA since it is represented by more countries and has the highest remaining budget after accounting for the super heavy-lift launch vehicle expenditures. The number of modules and rocket launches each government could contribute to the Vela mission is represented in Figure \ref{fig:optimization}. The mission plan incorporates programmatic redundancy in both the orbiting station modules and the launch vehicle capability to deliver those modules to Mars. 

\subsubsection{Limitations}
\label{sec:limitations}

This work focuses on direct costs associated with building and launching of vehicles and modules for supporting the proposed Vela human mission to Mars. It does not explicitly include the operational costs of the mission, such as the costs of resupply cargo launches or the costs associated with ground support. It also does not consider the overhead costs of managing the mission at the national level. The overall program budget from Figure \ref{fig:optimization}, however, is greater than the estimated fixed costs suggesting some margin to cover operational costs and development. The analysis also does not consider the research and development costs associated with the introduction of new technologies used for the mission like the development of new launch vehicles.
\subsection{Discussion}

\subsubsection{Econometric Analysis}

The econometric analysis presented in this work defined multiple model specifications to estimate the space budgets of the five space agencies. The models generated are a set of linear equations that predict the dependent variable (space budget) as a function of the independent variables (GDP per capita, researchers per million, military spending, education spending, and science R\&D spending). Due to the heteroscedasticity\footnote{Multiple combinations of independent variables capture a specific effect on the dependent variable.} of some\footnote{Heteroscedasticity was found in the data for China, the EU, Russia, and the USA.} of the data, multiple model specifications were created to obtain meaningful correlations between all independent variables and the dependent variable. 

Table \ref{tab:trends} summarizes the between-variable correlations from all the model specifications included in Appendix B. The trends, either positive or negative, are reported to show how the independent variables affect the space budget of each agency analyzed. The authors do not claim to have adequate knowledge to explain why all these correlations exist as they do, but rather the intent is to show that several complicated factors influence each space agencies budget, and that those factors can change or even have completely opposite effects from one agency to the next. The following discussion on potential hypothesis explaining the results is presented to give the reader a better ability to interpret the table and draw their own conclusions if desired.  

Starting at the bottom row of the table, GDP/Capita has a positive correlation with the space budget for all agencies, except JAXA where a negative value is observed with the caveat that the result was not statistically significant at the 5\% confidence interval. This generally means that an increase in GDP/Capita will result in an increase in each agency's space budget, which is perhaps consistent with expectations. Interestingly, the space budgets of China, Russia, and the USA exhibit a positive correlation with military spending, while the space budgets of the EU and Japan exhibit a negative correlation with military spending. This suggests a difference in how the governments funding these agencies view their utility. As an example, a positive correlation between military and space spending may indicate national attitudes of space capability being commensurate with military strength. A negative correlation between these variables may indicate prioritizing funding one program over the other based on current geopolitical events. Education spending shows a positive correlation with space spending for all agencies except NASA. This could be explained by the US political system, where constantly changing political parties have different beliefs of where to allocate funding. Spending in Science R\&D is typically a positive correlation with space budget, except for Roscosmos. Researchers in R\&D per million exhibited mixed correlation results across the agencies. CNSA and NASA have conflicting results from the different model specifications, JAXA and Roscosmos show a positive correlation, and ESA shows a negative correlation.

These results highlight the complex nature of space budget allocations for highly-funded space programs. Some of these budgets will increase in the future and some may decrease. This further supports the previous claims that a large amount of programmatic redundancy should be built into the first manned Mars mission to distribute costs, reinforce commitments, and lead to a successful mission outcome. 

\subsubsection{Mission Analysis}

Section 3 presents a cost-optimized space mission plan to establish a human presence on an orbiting space station at Mars. The number of modules and rocket launches were divided among the participating space agencies according to their existing and planned technological capabilities and funding allocations. The resulting Vela mission analysis includes an orbiting Mars space station with seven total modules requiring seven payload launches and one additional launch for the Vela crew. For a mission including station module development, rocket launches, a crew vehicle, and a service module, the total cost is approximately \$25B. The estimated total available budget for the Vela mission, assuming an annual contribution based on each agency's 2021 budget over a five year period, is \$34.3B. 

The mission analysis proposes a scenario for building a human space station at Mars using realistic budget and technological contributions while maintaining 27\% of the total Vela budget for operational and programmatic expenses. Although the true breakdown of modules, rocket launches, and funding levels will be dependent on the reality of budget allocations and technological capabilities at some future date, this study shows the feasibility of how such a mission could reasonably be organized under present-day constraints.

\subsection{Conclusions}

Using existing and planned capabilities and knowledge in human spaceflight, this work presented the Vela mission scenario for an orbiting space station at Mars. The work first conducted an econometric analysis to understand trends in space budget allocations for each of the five space agencies analyzed: CNSA, ESA, JAXA, Roscosmos, and NASA. Those findings were then leveraged to conduct a mission cost analysis showing viability of such an endeavor. 

Key findings from the econometric analysis of the space agencies' budgetary data are below.

\begin{enumerate}
    \item The effects of several independent variables, namely GDP/Capita, researchers per million people, military spending, education spending, and science R\&D spending, were correlated with each space program's budget allocations. 
    \item The results identify trends for each agency and highlight the need for programmatic redundancy to increase the odds of success for the first manned mission to Mars. 
\end{enumerate}

\noindent Conclusions from the mission analysis are reported below. 

\begin{enumerate}
    \item The capability to carry a crew to Mars is constrained to three countries which are currently developing super-heavy-lift launch vehicles: China, Russia, and the USA.
    \item Massive human habitat modules for a Martian space station each require a launch on a super heavy-lift launch vehicle. The cost of launching payload on a super heavy-lift launch vehicle for deep space exploration is expected to to remain the primary driver of total mission cost.
    \item The Vela mission analysis demonstrated the feasibility of establishing a human presence at Mars using realistic financial contributions (from current space funding levels) as well as both existing and planned technological capabilities of a team of international partners.
\end{enumerate}
\subsection{Future Work}

This work presented methods for modeling space agency funding levels and space exploration capabilities for China, the EU, JAXA, Roscosmos, and the USA. Future work should consider collaboration among a larger group of cooperating government space agencies in mission planning and econometric analysis, particularly as more space agencies develop deep space launch and exploration capabilities (e.g., India and the UAE). Future work should also consider developing a higher-fidelity Mars mission plan, incorporating scheduling constraints due to funding availability, operational and overhead costs, timeline constraints due to launch window availability, and the reusability of super heavy-lift launch vehicles. Lastly, the econometric analysis should be further leveraged to forecast future space agency budget allocations for realistic program funding levels over the duration of the mission. 
\subsection*{Acknowledgements}

The authors thank Professor Johannes Fedderke (School of International Affairs, The Pennsylvania State University) for his insight and guidance in performing econometric analysis.

\footnotesize
\renewcommand\bibsection{\subsection*{References}}
\bibliography{NatureAbrv,References}
\bibliographystyle{plainnat}

\normalsize
\subsection*{Authors}

\parpic{\includegraphics[width=1in,clip,keepaspectratio]{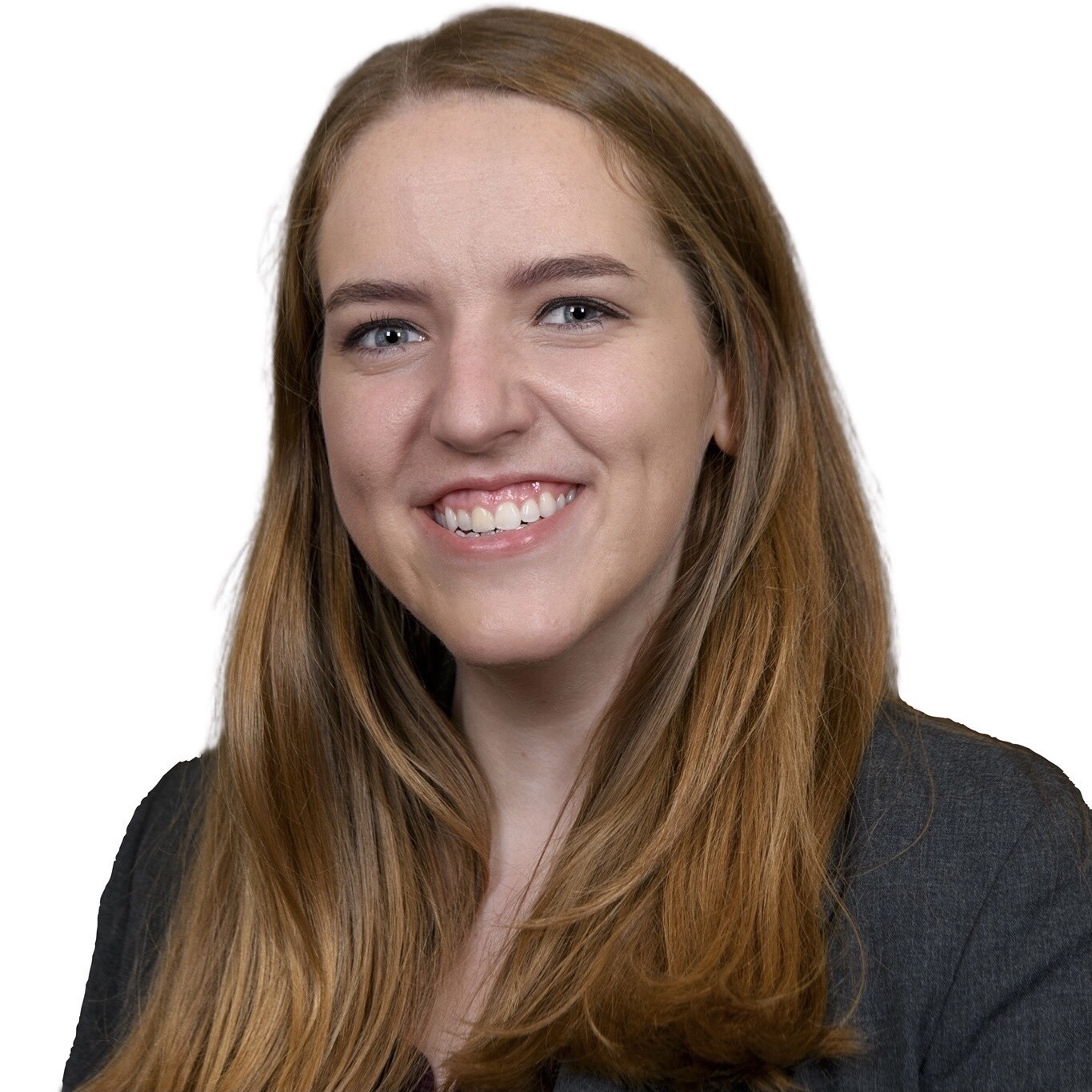}}
\noindent \textbf{Holly Dinkel} \href{https://www.linkedin.com/in/hollydinkel}{\faLinkedinSquare} \href{https://hollydinkel.github.io}{\faHome} is pursuing a Ph.D. in aerospace engineering at the University of Illinois Urbana-Champaign where she researches robotic caretaking as a NASA Space Technology Graduate Research Fellow with the NASA Ames Research Center Intelligent Robotics Group and the NASA Johnson Space Center Dexterous Robotics Laboratory.

\parpic{\includegraphics[width=1in,clip,keepaspectratio]{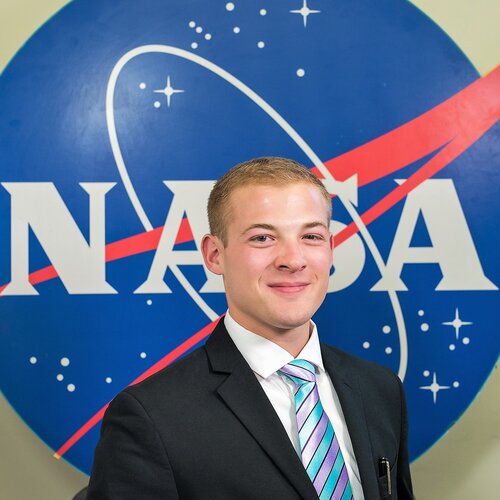}}
\noindent \textbf{Jason Cornelius} \href{https://www.linkedin.com/in/jason-k-cornelius}{\faLinkedinSquare} is pursuing a Ph.D. in aerospace at the Pennsylvania State University in University Park, PA where he supports NASA's Dragonfly New Frontiers mission as a NSF Graduate Research Fellow. He also works as an aerospace engineer in the Aeromechanics Office at the NASA Ames Research Center in Moffett Field, CA.

\end{multicols}
\normalsize
\newpage 
\subsection*{\centering Appendix A: Epilogue}

\justifying

\doublespacing
\textit{The nobility of the human spirit grows harder for me to believe in. War, zealotry, greed, malls, narcissism. I see a backhanded nobility in excessive, impractical outlays of cash prompted by nothing loftier than a species joining hands and saying ``I bet we can do this.'' Yes, the money could be better spent on Earth. But would it? Since when has money saved by government red-lining been spent on education and cancer research? It is always squandered. Let’s squander some on Mars. Let’s go out and play.}

\begin{flushright} --Mary Roach, \textit{Packing for Mars: The Curious Science of Life in the Void}\end{flushright}

\vspace{2.5em}

\doublespacing
In 2020-2021, the authors participated in the Stanford U.S.-Russia Forum, a year-long effort bringing together students from both countries to engender positive change in the relations between them. The results were presented at the 2021 IAF Global Space Exploration Conference in St. Petersburg, Russia. An obvious and prescient question was posed to the authors, ``Why limit the study to only Russia and the United States?'' This led the authors to conduct the present study including the five most funded space agencies, with nearly two dozen countries represented. The goal was to show what could realistically be accomplished if these countries found the will to work together to tackle the most audacious space mission of our generation: manned exploration of Mars. 

The authors chose Vela as the title of this paper, which means `sail' in Latin. Vela is the sail of the ship Argo in the constellation cluster Argo Navis. In Greek mythology, the Argonauts sailed on this ship in pursuit of the riches of their time. The authors consider the prospective of multi-national collaboration and programmatic redundancy on such a large mission to be the sail that will carry the project through to fruition. 
\newpage
\subsection*{\centering Appendix B: Johansen VECM Results}
\label{sec:appendixb}

\centering

\begin{flushleft}The Johansen VECM models for CNSA with three model specifications are shown in Tables \ref{tab:cnsa1}--\ref{tab:cnsa4}.\end{flushleft}

\begin{tablehere}
\captionsetup{type=table}
\caption{CNSA Model Specification 1, $\chi^2 = 295.4$}
\includegraphics[width=0.67\textwidth]{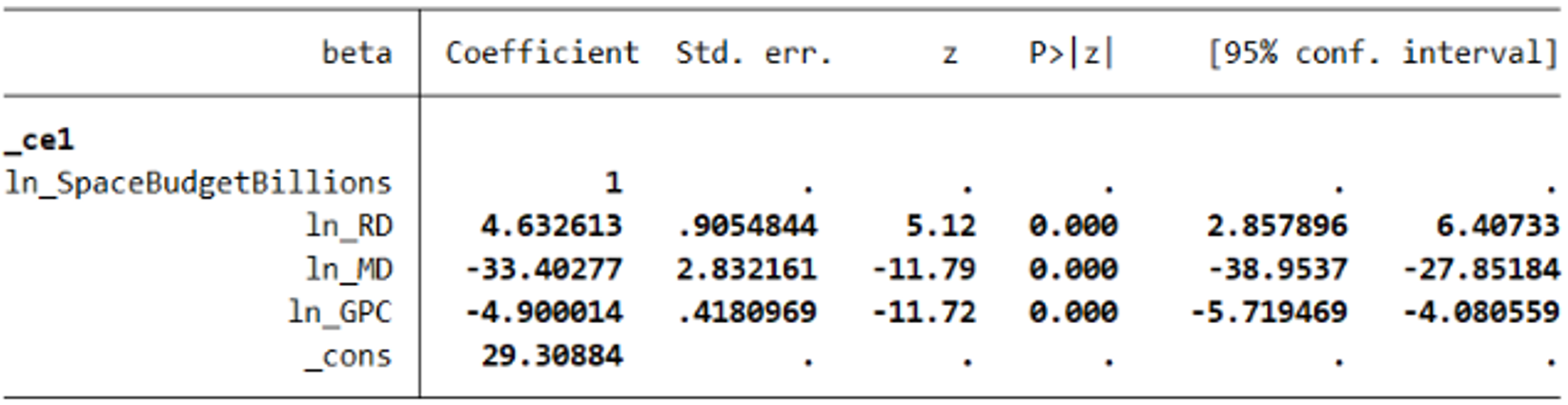}
\label{tab:cnsa1}
\end{tablehere}

\begin{tablehere}
\captionsetup{type=table}
\caption{CNSA Model Specification 2, $\chi^2 = 119.1$}
\includegraphics[width=0.67\textwidth]{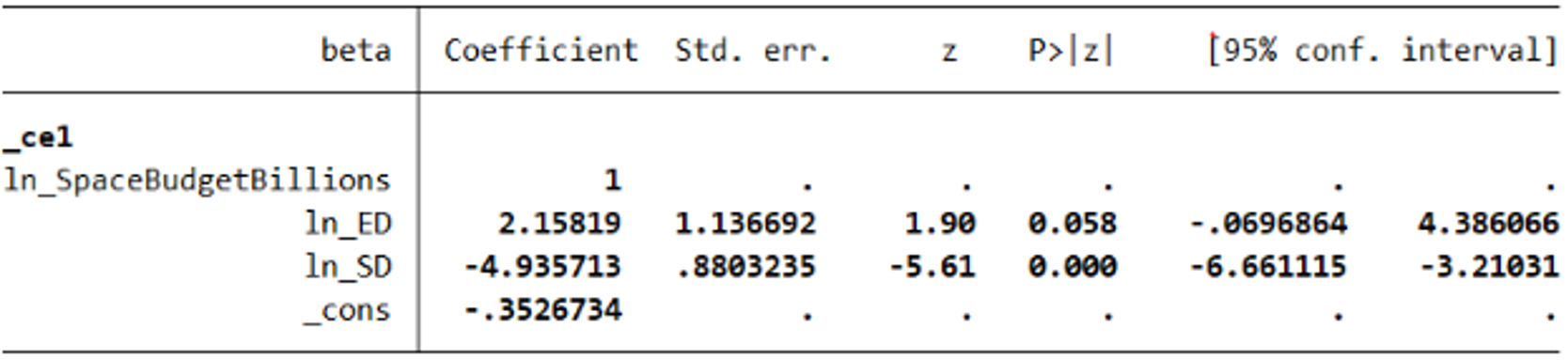}
\label{tab:cnsa2}
\end{tablehere}

\begin{tablehere}
\captionsetup{type=table}
\caption{CNSA Model Specification 3, $\chi^2 = 119.5$}
\includegraphics[width=0.67\textwidth]{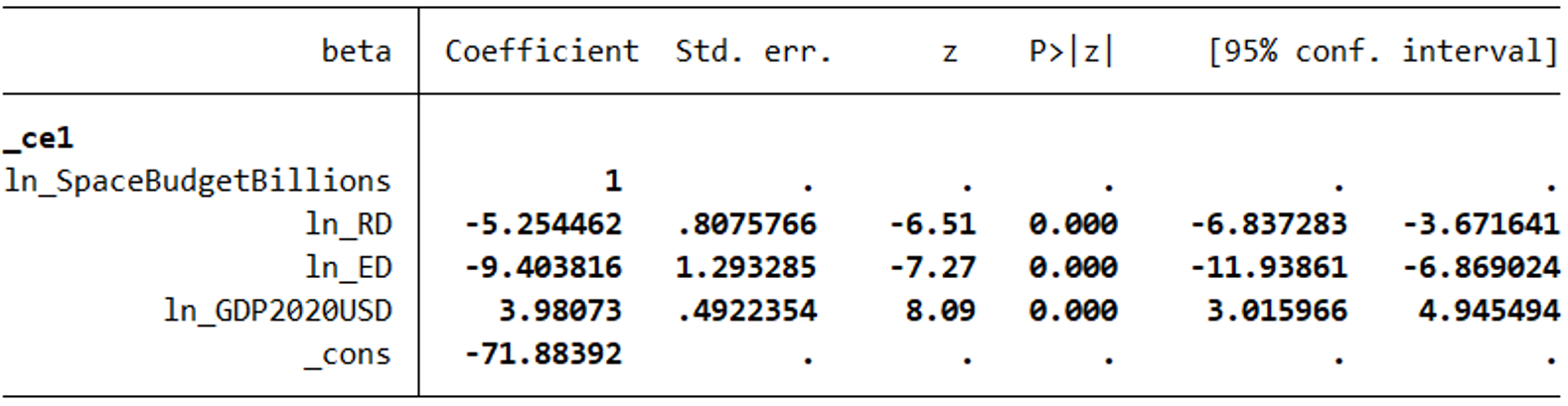}
\label{tab:cnsa4}
\end{tablehere}

\begin{flushleft} The Johansen VECM models for ESA with two model specifications are shown in Tables \ref{tab:esa1}--\ref{tab:esa2}. \end{flushleft}

\begin{tablehere}
\captionsetup{type=table}
\caption{ESA Model Specification 1, $\chi^2 = 7860.4$}
\includegraphics[width=0.67\textwidth]{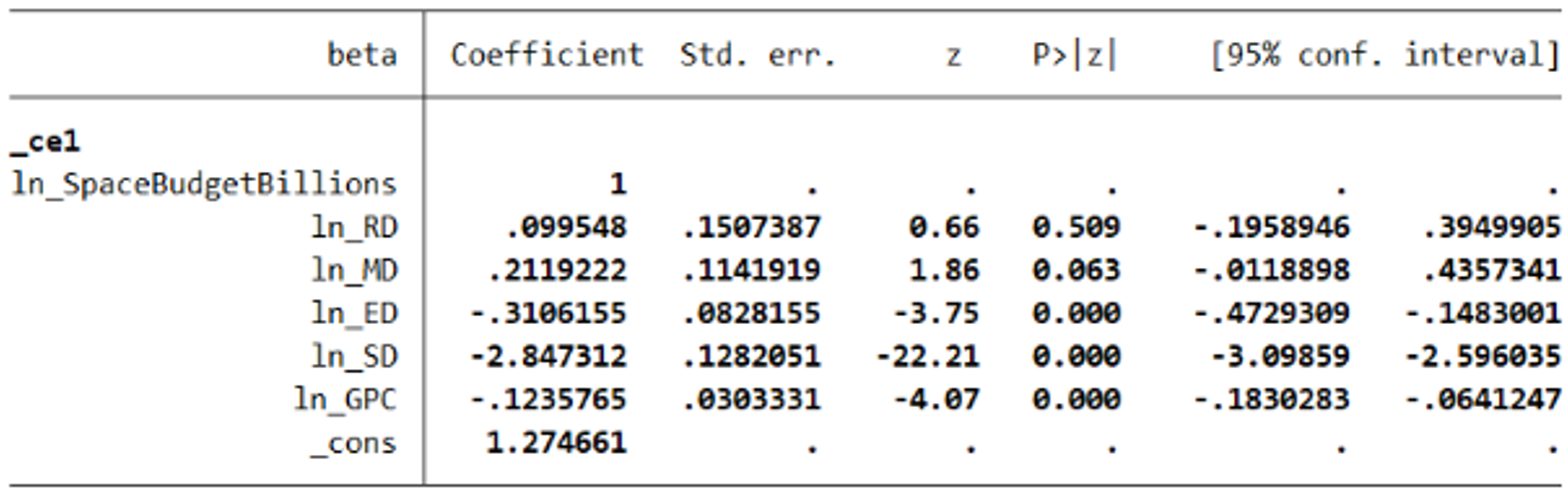}
\label{tab:esa1}
\end{tablehere}

\newpage
\begin{tablehere}
\captionsetup{type=table}
\caption{ESA Model Specification 2, $\chi^2 = 5233.3$}
\includegraphics[width=0.67\textwidth]{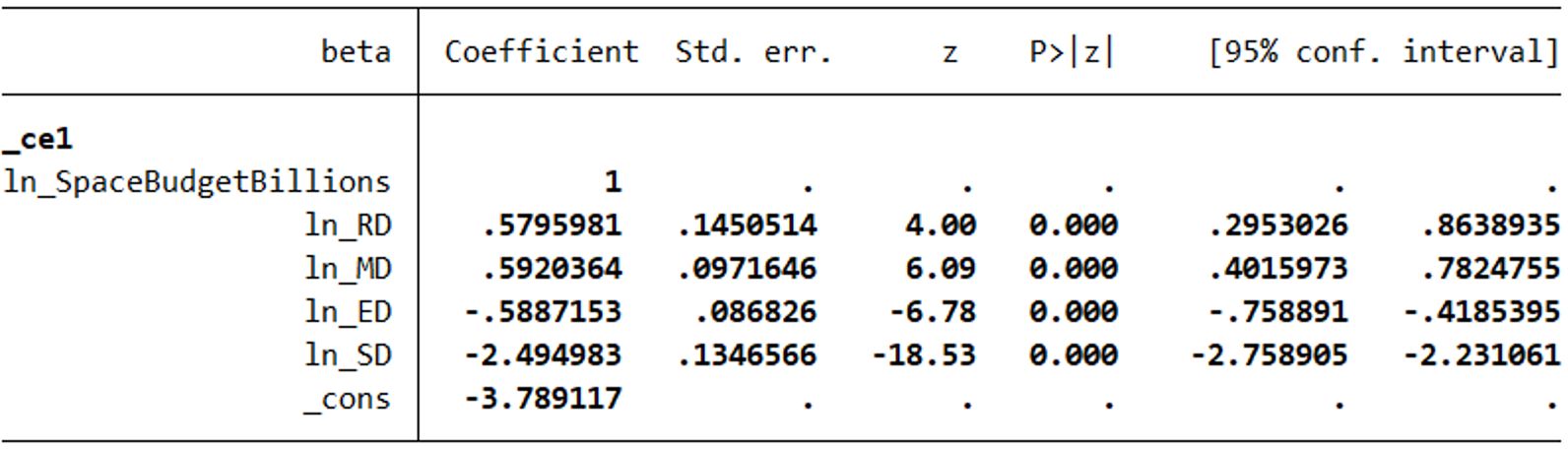}
\label{tab:esa2}
\end{tablehere}

\begin{flushleft} The Johansen VECM model for JAXA with one model specification is shown in Table \ref{tab:jaxa1}. \end{flushleft}

\begin{tablehere}
\captionsetup{type=table}
\caption{JAXA Model Specification$\chi^2 = 148.2$}
\includegraphics[width=0.67\textwidth]{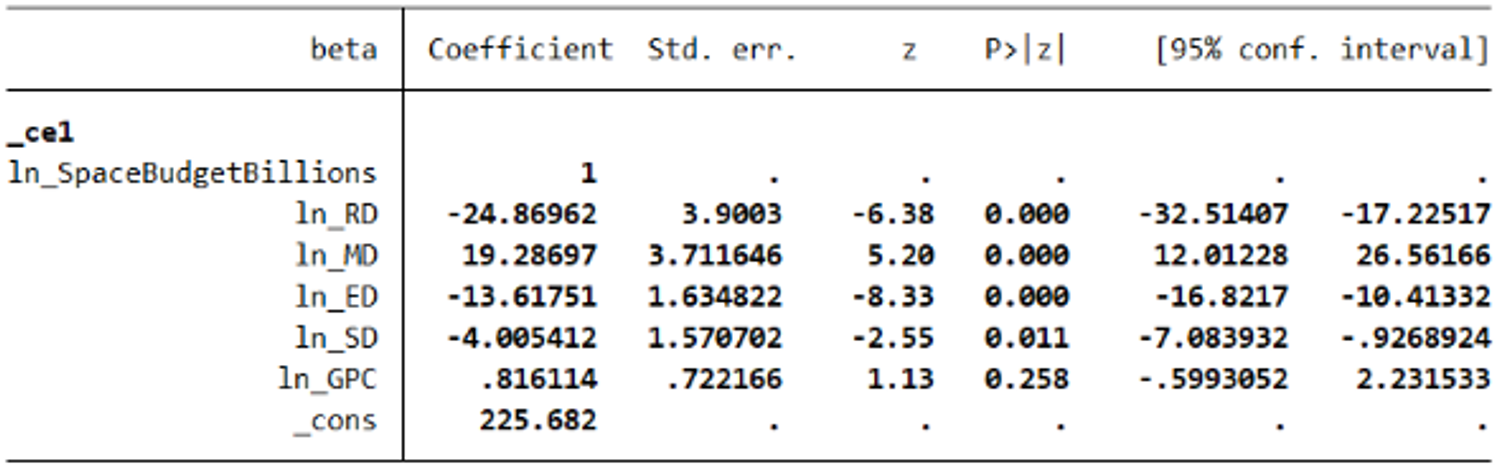}
\label{tab:jaxa1}
\end{tablehere}

\begin{flushleft} The Johansen VECM models for Roscosmos with three model specifications are shown in Tables \ref{tab:ros1}--\ref{tab:ros4}. \end{flushleft}

\begin{tablehere}
\captionsetup{type=table}
\caption{Roscosmos Model Specification 1, $\chi^2 = 60.4$}
\includegraphics[width=0.67\textwidth]{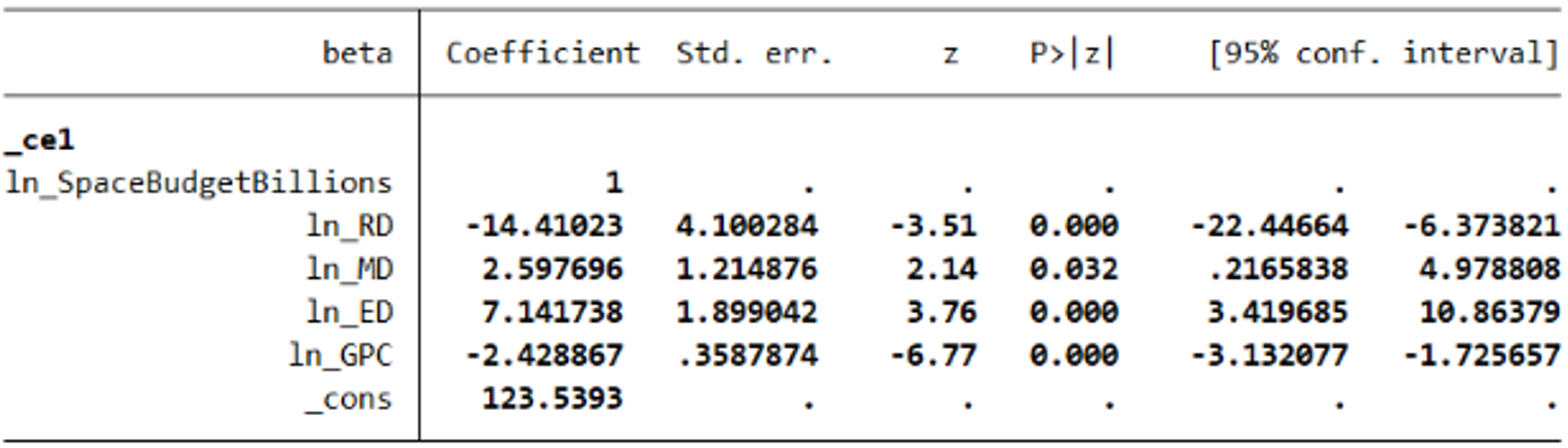}
\label{tab:ros1}
\end{tablehere}

\begin{tablehere}
\captionsetup{type=table}
\caption{Roscosmos Model Specification 2, $\chi^2 = 43.1$}
\includegraphics[width=0.67\textwidth]{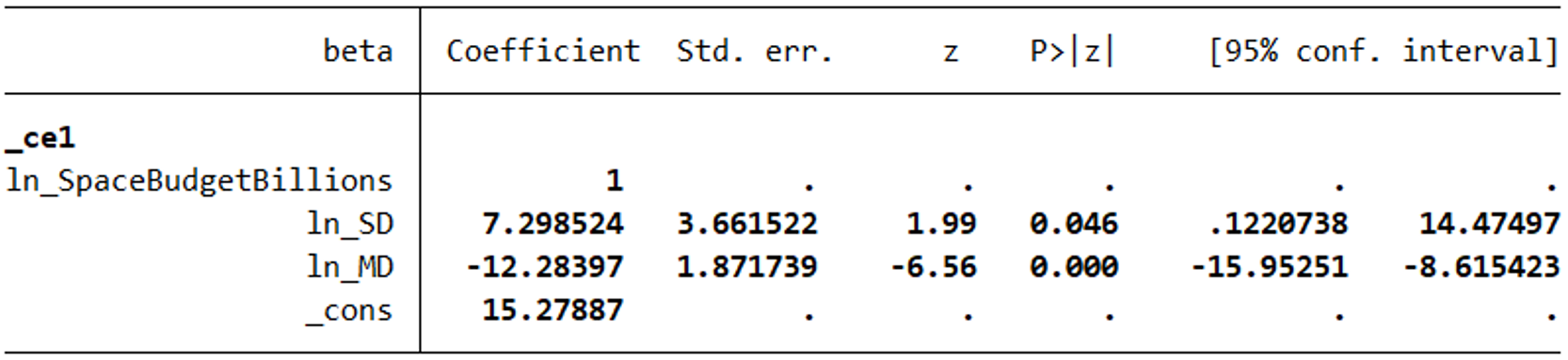}
\label{tab:ros2}
\end{tablehere}

\newpage
\begin{tablehere}
\captionsetup{type=table}
\caption{Roscosmos Model Specification 3, $\chi^2 = 87.0$}
\includegraphics[width=0.67\textwidth]{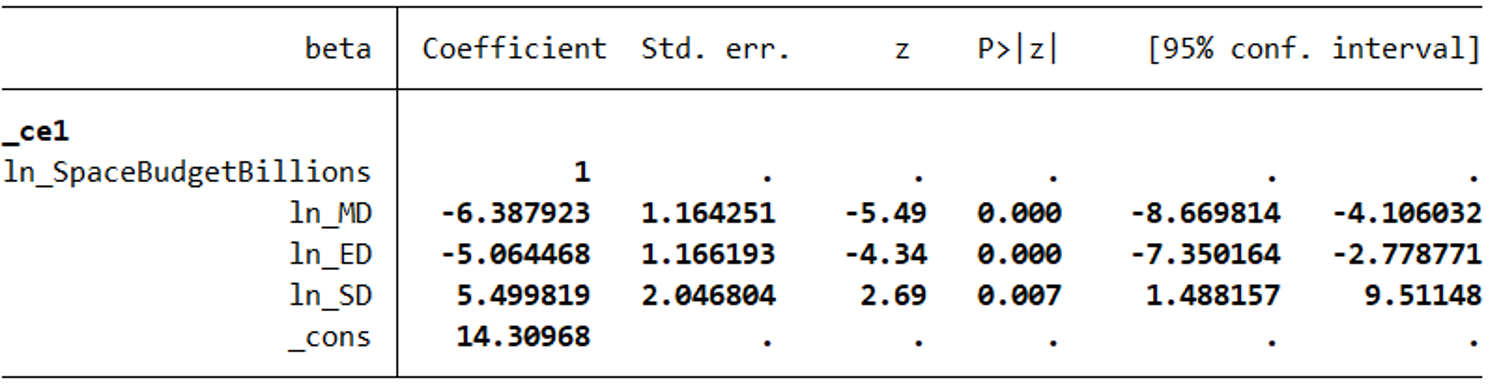}
\label{tab:ros4}
\end{tablehere}

\begin{flushleft} The Johansen VECM models for NASA with three model specifications are shown in Tables \ref{tab:nasa1}--\ref{tab:nasa3}. \end{flushleft}

\begin{tablehere}
\captionsetup{type=table}
\caption{NASA Model Specification 1, $\chi^2 = 565.2$}
\includegraphics[width=0.67\textwidth]{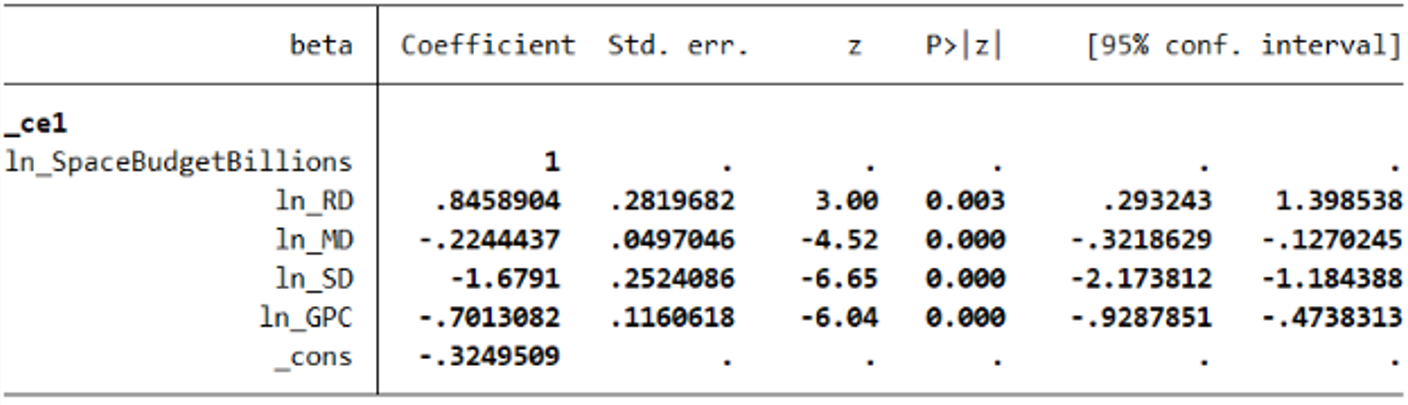}
\label{tab:nasa1}
\end{tablehere}

\begin{tablehere}
\captionsetup{type=table}
\caption{NASA Model Specification 2, $\chi^2 = 141.2$}
\includegraphics[width=0.67\textwidth]{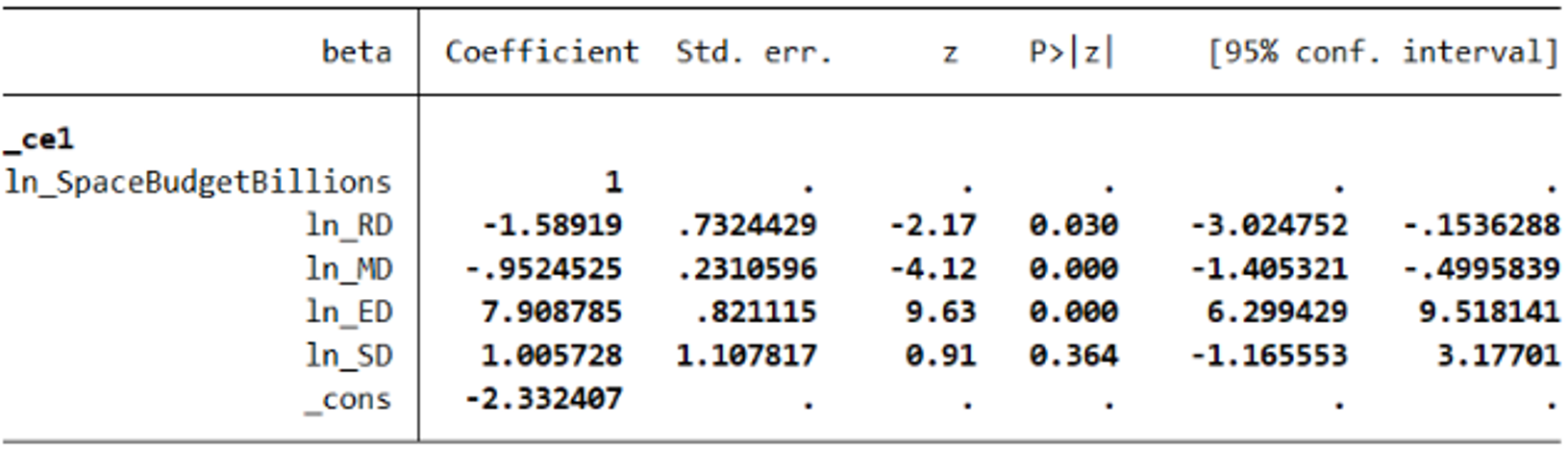}
\label{tab:nasa2}
\end{tablehere}

\begin{tablehere}
\captionsetup{type=table}
\caption{NASA Model Specification 3, $\chi^2 = 270.6$}
\includegraphics[width=0.67\textwidth]{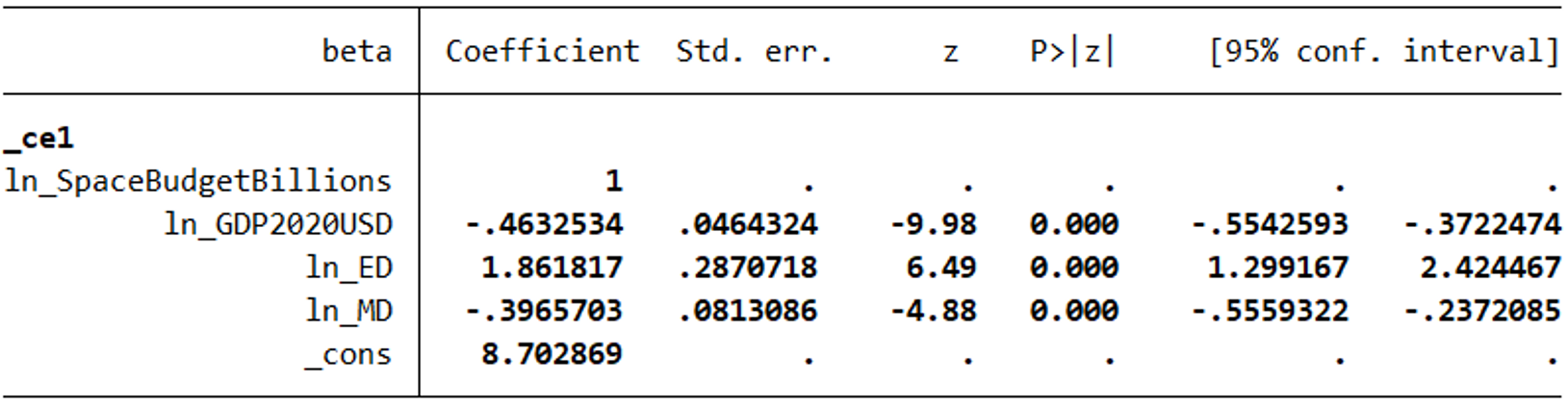}
\label{tab:nasa3}
\end{tablehere}

\end{document}